\documentclass[12pt,preprint]{aastex}

\newcommand{\pref}{\protect\ref}
\newcommand{\solrad}{\ifmmode{R}_{\rm S}\else${R}_{\rm S}$\fi}
\newcommand{\solmas}{\ifmmode{M}_{\rm S}\else${M}_{\rm S}$\fi}

\newcommand{\ctn}{\ifmmode\kappa\else$\kappa$\fi}

\newcommand{\term}[2]{\mbox{$\,^{#1}{\rm #2}$}}
\def\term#1 #2/{\mbox{$\,^{#1}{\rm #2}$}}

\newcommand\lta { \mathrel {\hbox to 0pt {\lower 3.7pt \hbox{$\sim$}
      \hss} \raise 1.7pt \hbox{$<$}}}
\newcommand\gta { \mathrel {\hbox to 0pt {\lower 3.7pt \hbox{$\sim$}
      \hss} \raise 1.7pt \hbox{$>$}}}
\newcommand{\philemail}{judge@ucar.edu}

\newcommand{\width}{w}
\newcommand{\widi}{v_i}
\newcommand{\delt}{\tau}
\newcommand{\gauss}[3]{\ensuremath e ^{ -\left ( {{#1-#2}\over #3} \right )^2 }}

\newcommand{\figone}{
\begin{figure}[] 
\epsscale{1.0}
\plotone{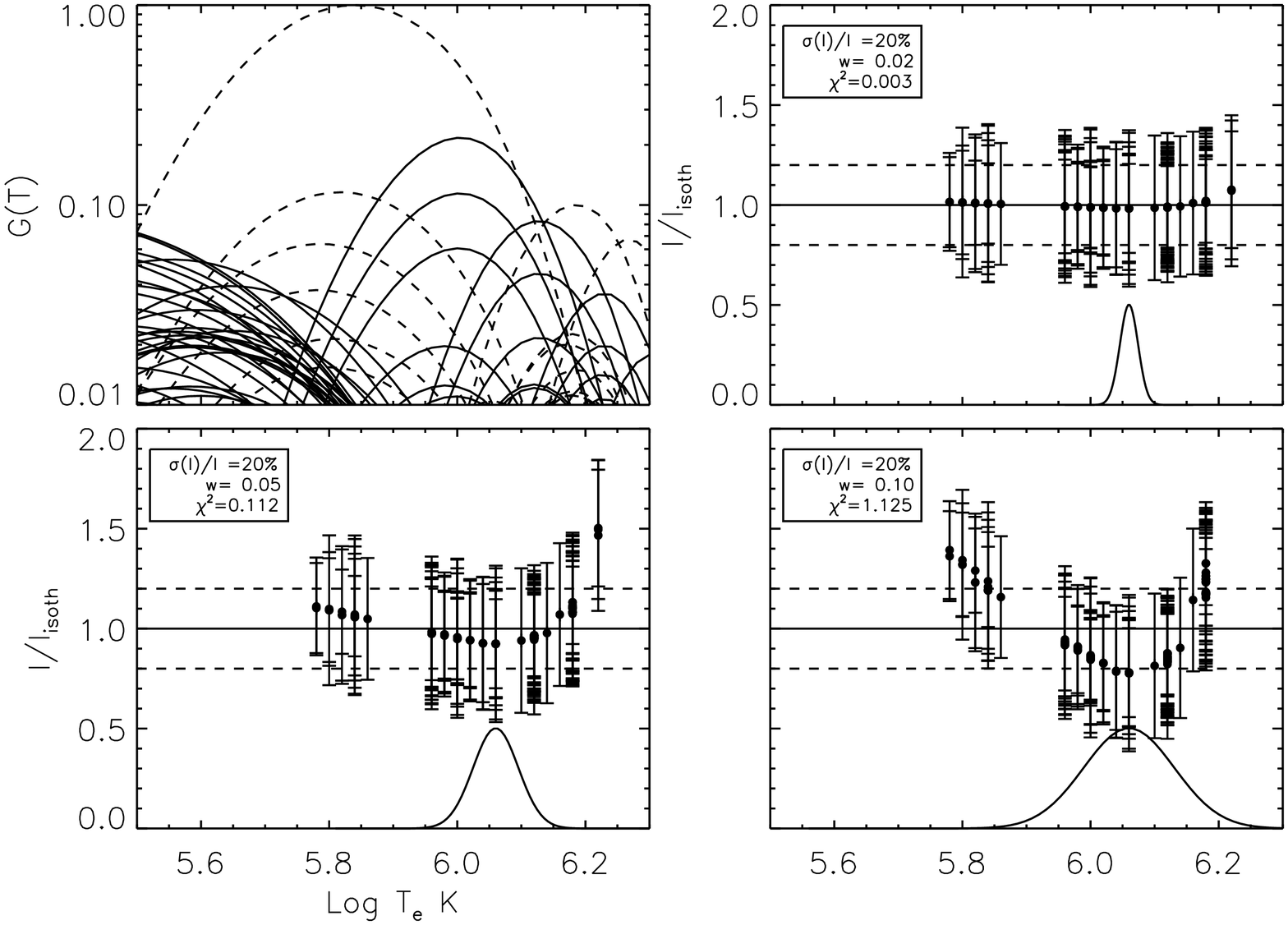}  
\caption{\label{fig:iron}
$G(T)$ functions for Fe VIII-Fe XVII are shown as a function of $T$ (logarithm of the electron temperature) in the 
upper left panel.  Solid and dashed lines mark ions with odd and even charges repectively, for clarity.  The other panels plot the ratios of intensities computed using the Gaussian functions shown (solid lines) divided by the intensities for an isothermal calculation centered at the same 
temperature (in this case $T=$log$_{10}1.2\times10^6$ K).  The uncertainties are set to $\gta 20\%$ of the 
computed line intensities (see text) and the computed $\chi^2$ parameters are listed with the Gaussian width $\width$.
}
\end{figure}
}

\shortauthors{Judge}
\shorttitle{}

\slugcomment{}

\begin{document}

%###############################################################################
%
%     OPENING
%
%###############################################################################

\title{Coronal emission lines as thermometers}
\author{Philip G. Judge}
\affil{High Altitude Observatory,
       National Center for Atmospheric Research\footnote{The National %
       Center for Atmospheric Research is sponsored by the %
       National Science Foundation},
       P.O.~Box 3000, Boulder CO~80307-3000, USA; \philemail}
%\date{Version \today}

%###############################################################################
%
%     ABSTRACT
%
%###############################################################################

\begin{abstract}

Coronal emission line intensities are commonly used to measure electron temperatures using emission measure and/or line ratio methods.  In the presence of systematic errors in atomic excitation calculations and data noise, the information on underlying temperature distributions is fundamentally limited. Increasing the number of emission lines used does not necessarily improve the ability to discriminate between different kinds of temperature distributions.

\end{abstract}


\section{Introduction}

Spectroscopic measurements of the temperature of coronal plasma have been made 
for decades \citep[e.g.][]{Seaton1962c,Noci2003}.  
\citet{Feldman+others1999,Feldman+Landi2008} and \citet{Landi+Feldman2008} 
presented intriguing  evidence that electron temperatures of 
solar coronal plasma, measured from emission line spectra obtained high in the coroma, are clumped into several peaks, 
and are not broadly distributed.   Narrow distributions of plasma temperature
have important implications for the energy balance of the corona.  
In the methods used, 
frequency integrated line intensities $I$ are assumed to be simple functions $G(T)$ of the logarithm of the 
electron temperature, $T$, because of the dominance of two body collisional processes
which lead to the well-known ``coronal approximation'' \citep[e.g.][]{Woolley+Allen1948,Seaton1964a}.   Two approaches were 
used.  In one they solved for the differential emission measure $\xi(T)$ which is an optimal solution to the inverse problem:
\begin{equation}
I_i = \int G_i(T) \xi(T) dT,\ \ \ \ i=1\ldots n
\end{equation}
where there is a set of $n$ different emission lines.  The other method sought the single logarithmic temperature $T_0$ 
such that $\xi(T) \propto \delta(T-T_0)$, by plotting $I_i/G_i(T)$ and identifying intersection 
points for the observed lines.  These approaches are,
in fact, formally equivalent \citep{McIntosh+Brown+Judge1997}.
In both cases, information must be added to obtain both $\xi(T)$ and $T_0$: in the first case the problem has to be ``regularized'' as it is ill-posed \citep[e.g.][]{Craig+Brown1986}, one searches for example for the  ``least structured'' solution $\xi(T)$ that is compatible with the data.
In the second case it is assumed that the plasma is indeed approximately isothermal.  
In the analysis of \citet{Feldman+Landi2008}, the individual peaks in the $\xi(T)$ functions have widths of 0.1-0.2 in $T$, and the curves of $I_i/G_i(T)$ intersect one another within similar margins. 

Some general questions arise.  Given typical uncertainties in observable and model parameters, 
are the observed
data compatible with different $\xi(T)$ distributions?  What, then, is the accuracy of the derived temperatures?   In this paper these questions are addressed 
by asking, how broad can $\xi(T)$ functions 
be to be incompatible with observed data. By how much can one change the temperature of an
isothermal plasma before the differences in line intensities become significant? 
In the problem at hand there are unavoidably large and systematic uncertainties in the $G(T)$ functions. These uncertainties limit the information which can be extracted from emission line spectra.  
It is found that acceptable widths $\width$ of the $\xi(T)$ functions exceed the precision by which the lines can in principle determine that two different plasmas have slightly different temperatures.  Thus, these widths set the lower limit to the ability of emission line techniques to diagnose electron temperatures.

\section{Calculations of temperature sensitive lines}

\subsection{Uncertainties in $G(T)$}

$G(T)$ functions depend linearly on electron impact excitation rates, on
elemental abundances, and on factors influencing the ionization fraction for a 
given line.  
For coronal ions, uncertainties in electron impact excitation cross sections, based primarily upon sophisticated calculations, are typically 
$\pm10$\%  but can be considerably higher \citep[e.g.][]{Storey+Zeippen+leDourneuf2002}. 
Other uncertainties in $G(T)$ arise from element abundance uncertainties,  bound-free cross sections (ionization, recombination),
and can include systematic errors from non-ionization equilibrium effects, from any 
significant 
non-thermal populations of electrons, modifications of dielectronic recombination rates due to finite density plasmas \citep[e.g.][]{Summers1974a,Badnell+others2003,Judge2007a}, and even radiative transfer.  
Given these considerations, we adopt fractional uncertainties of 
$\epsilon=0.2$ for all $G(T)$ functions, which 
should probably be considered a lower limit.

\subsection{Numerical calculations}

Calculations were performed for lines of Fe VIII - Fe XVII  using the {\em DIPER} package \citep{Judge2007a}.  The upper left panel of Figure~\pref{fig:iron} shows $G(T)$ functions computed for a typical coronal electron pressure of 0.1 dyne cm$^{-2}$.  In constructing this figure, intensities $I_{isoth}$ were computed using an isothermal plasma at 
$1.3\times10^6$K.  Intensities were then recomputed using a 
Gaussian function with unit ($-\infty \rightarrow +\infty$) integral:
\begin{equation}
  \label{eq:1}
  \xi(T)=  \frac{1}{\sqrt{\pi}\width}\gauss{T}{T_0}{\width}
\end{equation}
The intensities were recomputed, and a value of $\chi^2$ computed using 
the above uncertainties in $G(T)$ added in quadrature to 
the uncertainty from photon counting statistics, assuming that the brightest line in the spectrum has accumulated $10^4$ counts.  These uncertainties are not unrealistic values for what might be expected from UV and EUV spectrographs of the modern era.
This particular calculation shows that an isothermal plasma 
with $T=T_0$
cannot {\em formally}
be distinguished from a plasma with Gaussian distributions of width $\width$ less than 0.1.  {\em Informally},  the figure shows a systematic increase in the ratio $I/I_{isoth}$ in lines with $G(T)$ peaking away from $T_0$.  One may be tempted to claim that the curvature indicates the presence of broader distributions.
But these calculations are made with the {\em same} set of 
$G(T)$ functions, where in reality one would compare these theoretical values 
with those (unknown) functions, free of systematic errors, present in the real Sun. 

\subsection{Analytical approximations}

To generalize these results, 
analytical results can be obtained using suitably simple functions to approximate 
$G_i(T)$. 
Here, Gaussian functions are adopted both for $\xi(T)$, as in eq.~(\pref{eq:1}),  and $G_i(T)$:
\begin{equation}
  \label{eq:2}
G_i(T)=a_i\gauss {T}{T_i}{\widi}. 
\end{equation}
Very similar results are obtained using 
$\xi(T)$ represented by a ``top-hat'' function of full width $2\width$.  These were chosen because they are not qualitatively dissimilar from the shapes of typical lines, 
and they are analytically simple.
Then 
\begin{eqnarray}
I_i &=& \frac{a_i}{\sqrt{\pi}\width}  \int_{-\infty}^{\infty}  
\gauss{T}{T_i}{\widi} \gauss{T}{T_0}{\width}  dT,\ \ \ \ i=1\ldots n\\
    &=& \frac{a_i \widi}{\sqrt{\width^2+\widi^2} } \ e^{ -\frac{ (T_0-T_i)^2}{\width^2+\widi^2}}
\end{eqnarray}
When $\width \rightarrow 0$, then 
$
I_i \equiv I_i^0= {a_i}  \gauss{T_0}{T_i}{\widi}
$, i.e. the isothermal case.
If $\width^2 \ll \widi^2$ and for source temperatures $T_1=T_0+\delt$  ($\delt \ll |T_0-T_i|$- large values of $|T_0-T_i|$ are needed for significant temperature sensitivity between at least some lines),
these expressions can be expanded as a Taylor series, to yield changes in $I_I$ as functions of small values of $\width$ and $\delta$.   The first derivatives are
\begin{eqnarray} \label{third}
\frac{\partial I_i}{\partial T_0}  &=& 2 I_i^0  \ \frac{T_i-T_0}{\width^2+\widi^2} \\
\frac{\partial I_i}{\partial \width}  &=& 2 I_i^0 \ \width
\frac{1}{\width^2+\widi^2}  \left ( \frac{(T_i-T_0)^2}{\width^2+\widi^2} 
-\frac{1}{2} \right ).
\end{eqnarray}
Dividing by $I_i$, the first order fractional changes in $I_i$ resulting from
small changes $\delt$ and $\width$ are:
\begin{eqnarray} \label{third}
\Delta_T=\frac{1}{I^0_i}\frac{\partial I_i}{\partial T_0}  \delt &\approx& 2 \ \frac{T_i-T_0}{\widi^2} \delt \\
\Delta_\width = \frac{1}{I^0_i}\frac{\partial I_i}{\partial \width}  \width&\approx& 
\frac{2 \width^2}{\widi^2}  \left ( \frac{(T_i-T_0)^2}{\widi^2} 
-\frac{1}{2} \right )
\end{eqnarray}
%
%This equation shows that the intensity of line $i$, when arising from a plasma whose
%temperature distribution has width $2\width$,  has a fractional correction of
%$\frac{\width^2}{3\widi^2} \left [ 2x_i^2-1\right ] $ compared with a plasma %isothermal at temperature $T_0$.   
These correction terms can be used to compute the 
reduced $\chi^2$ statistic, to first order, using the uncertainties $\epsilon_i$:
\begin{equation}
  \label{eq:chi2}
  \chi^2 = \frac{1}{n-1}\sum_i^n  \frac{(I_i - I_i^0)^2}{I_i^0\epsilon_i^2} 
             =  \frac{1}{n-1}\sum_i^n  \frac{\Delta_i^2}{\epsilon_i^2 }
\end{equation}
Define $x_i= |T_0-T_i| / \widi$, which measures the difference in temperature of the peak of the source relative to the peak of line $i$ in units of the Gaussian width of 
$G_(T)$.  Any sensible set of lines must include values of $x_i$ varying from 
0 to numbers $\gg 1$.  But for $x_i \gta 3$, the lines become very weak with 
accompanying large observational uncertainties.  Practically speaking, 
we can consider most
values of $x_i$ to be spread between  0 and 3. 
Any given set of lines
cannot be used to discriminate an isothermal source at  temperature $T_0$ from another 
at temperature $T_0\pm\delt$ when (using $\Delta=\Delta_T$), the majority of lines satisfy 
\begin{equation} \label{eqn:descrim}
|\delt| \lta \frac{\epsilon_i}{2x_i} \ \widi, \ \forall \   i=1\ldots n.
\end{equation}
Similarly one cannot discriminate an isothermal source at  temperature $T_0$ from a 
broadened distribution with Gaussian parameter $\delt$ when (using 
$\Delta=\Delta_\width$)
\begin{equation} \label{eqn:descrim}
\width \lta \sqrt{\frac{\epsilon_i}{|2x_i^2-1|}} \ \widi, \ \forall \   i=1\ldots n.
\end{equation}
Using $x_i \sim 1$, $\widi \sim 0.15$, $\epsilon \sim 0.2$, the set of emission lines lines has sensitivities to the peak of the temperature and its width given by $\width$ of 
\begin{eqnarray}
  \label{eq:3}
  \tau \sim 0.015, \ \ \ \ \ \ \ \width \sim 0.13.
\end{eqnarray}
This result for $\width$ is in reasonable agreement with the numerical
calculation. 
 The ``bluntness of the thermometer'' is, in this approximation, measured by 
$\width$. This quantity therefore determines the {\em accuracy} of temperature measurements.  The accuracy is a factor of 8 worse than the {\em precision} with which different isothermal plasma temperatures can be discerned, relative to one another, which is represented by $\tau$.
The width $\width$ 
is weakly dependent on the intrinsic uncertainties $\epsilon_i$, linearly dependent on widths $\widi$ of the $G_i(T)$ function, and inversely proportional to 
$|x_i|$.  

Given the weakness of lines with large $x_i$ values, the nature of the uncertainties $\epsilon_i$, it relatively little can be done to increase 
the thermometer's accuracy.  However, one can try to include very strong lines 
(large $a_i$) which are formed far from their peak temperature $T_i$, i.e. when $x_i > 3$, say, such that the counts are still sufficiently large to make $\epsilon_i$ approach the systematic uncertainties in atomic calculations.  One example might be the strong line of Fe~IX near 17.1 nm whose $G(T)$ function is prominent in Figure~\pref{fig:iron}.  In such cases 
\begin{equation}  \label{eqn:bigx}
\width \propto \widi \sqrt{\epsilon_i} / x_i\ \ \ \  \ \ (x_i \gg 1 )
\end{equation}
Note however, that such lines tend also to have large widths $\widi$.

Temperature estimates cannot however be improved simply by analyzing more lines with similar values of $x_i$, because then the largest acceptable width $\width$ is almost independent of the number of lines $n$ used.  The uncertainties in $\epsilon_i$ are, in an important sense, irreducible, as, even in the limit of perfect observations, they result from uncertainties in atomic cross sections and in our lack of knowledge of the conditions controlling level populations in the solar coronal plasmas.  

\section{ Discussion}

The largest values of $\width$ compatible with observations yield the accuracy 
with which emission lines can measure electron temperatures.  The widths of the distributions compatible with known sources of uncertainties set a natural limit on the sharpness of a detectable peak in the underlying emission measure distributions. 
Typically the
full width, $2\width$, is 0.2 to 0.3 in the logarithmic electron temperature.   The precision is an order of magnitude better, being estimated using the sensitivity of the spectra to two 
isothermal plasmas differing in logarithmic temperature by $\tau$.  
Although  this implies that two strictly isothermal plasmas {\em could} be 
differentiated by using line ratios to a precision of order $\epsilon_i \widi/2 \sim 0.015$ in the logarithmic electron temperature, we will perhaps never know if such isothermal plasmas exist with a width narrower than $\width$, based upon these data.

These properties prompt the following comments:

\begin{enumerate}
\item Error bars quoted below $\sim0.1$ in logarithmic temperatures are not credible using these techniques. Landi and Feldman quote error bars of 0.04 and 0.05 in their work which seeks to provide spectroscopic evidence that the coronal temperature is ``quantized''.
\item The widths (FWHM) of peaks in the $\xi(T)$ functions found by 
\citet[][their figure 5]{Landi+Feldman2008} are between 0.1 and 0.2. These widths are characteristic of the limit with which emission lines can determine the isothermality of the emitting plasma, and are therefore determined more by regularization than by the data themselves. Their analysis is consistent with isothermal plasma, but is limited by the uncertainties discussed here.
\item  Coronal temperatures are controlled by (unknown) heating mechanisms and cooling by 
heat conduction, flows, radiation.  The scalings of \citet{Rosner+Tucker+Vaiana1978}, based on simple energy balance considerations,  show that a given coronal electron temperature $T_e$ in a loop of length $L$ requires an energy flux density 
$$
{\cal F} \propto (T_eL)^{7/2}
$$
to sustain it agains conductive and radiation losses.  Therefore, for a given $L$, if $T_e$ is uncertain to $\pm0.15$ in its logarithm, the absolute value of log$_{10}{\cal F}$ can be determined from a given length loop to an accuracy of $\sim \pm0.5$.  The energy flux is constrained only to somewhere within a range spanning a factor of ten using emission lines. Conversely, variations of the energy flux ${\cal F}$ dissipated in coronal loops by $\lta 0.5$ in the logarithm cannot be distinguished through temperature dependent emission line methods.  
\item  {\em Relative} changes in ${\cal F}$ could, however, be detected to within about $\pm 12\%$, but again this precision depends on the plasma being nearly isothermal (i.e. actual $\width \sim \delt$), which is neither known from the data nor expected from
first principles.
\end{enumerate}

The analytical results offer some hope that carefully selected sets of emission lines can be used to reduce the widths of distributions compatible with the data.  Specifically, using lines within the same element removes errors arising from incorrect abundances, and can include lines with large values of $|x_i|$.  In this case just a few lines with different values of $x_i$ can reduce $\width$ via equation~(\pref{eqn:bigx})).  Using lines from within the same ion stage yields significantly  smaller values of $\epsilon_i \sim 0.1$, but while the precision improves linearly with $\epsilon_i$
the accuracy improves only quadratically. 

\acknowledgments
The author is grateful to Scott McIntosh for carefully reading the manuscript.

\figone

%\figtwo

\end{document}